\documentclass[reprint,nofootinbib,superscriptaddress,amsmath,amssymb,aps,pra,floatfix]{revtex4-1}
\usepackage{graphicx,color}
\usepackage{bm}
\usepackage{hyperref}
\def\vec#1{\bm{#1}}
\def\abs#1{\left\lvert#1\right\rvert}

\def\ef{\varepsilon_F}

\begin{document}

\title{Scattering of two heavy Fermi polarons: resonances and
  quasibound states}
\author{Tilman Enss}
\affiliation{Institut f\"ur Theoretische Physik,
  Universit\"at Heidelberg, 69120 Heidelberg, Germany}
\author{Binh Tran}
\author{Michael Rautenberg}
\author{Manuel Gerken}
\author{Eleonora Lippi}
\affiliation{Physikalisches Institut,
  Universit\"at Heidelberg, 69120 Heidelberg, Germany}
\author{Moritz Drescher}
\affiliation{Institut f\"ur Theoretische Physik,
  Universit\"at Heidelberg, 69120 Heidelberg, Germany}
\author{Bing Zhu}
\affiliation{Physikalisches Institut,
  Universit\"at Heidelberg, 69120 Heidelberg, Germany}
\affiliation{Hefei National Laboratory for Physical Sciences at the
  Microscale and Department of Modern Physics, and CAS Center for
  Excellence and Synergetic Innovation Center in Quantum Information
  and Quantum Physics, University of Science and Technology of China,
  Hefei 230026, China}
\author{Matthias Weidem\"uller}
\affiliation{Physikalisches Institut,
  Universit\"at Heidelberg, 69120 Heidelberg, Germany}
\author{Manfred Salmhofer}
\affiliation{Institut f\"ur Theoretische Physik,
  Universit\"at Heidelberg, 69120 Heidelberg, Germany}
\date{\today}

\begin{abstract}
  Impurities in a Fermi sea, or Fermi polarons, experience a Casimir
  interaction induced by quantum fluctuations of the medium.  When
  there is short-range attraction between impurities and fermions,
  also the induced interaction between two impurities is strongly
  attractive at short distance and oscillates in space for larger
  distances.  We theoretically investigate the scattering properties
  and compute the scattering phase shifts and scattering lengths
  between two heavy impurities in an ideal Fermi gas at zero
  temperature.  While the induced interaction between impurities is
  weakly attractive for weak impurity-medium interactions, we find
  that impurities strongly and attractively interacting with the
  medium exhibit resonances in the induced scattering with a sign
  change of the induced scattering length and even strong repulsion.
  These resonances occur whenever a three-body Efimov bound state
  appears at the continuum threshold.  At energies above the continuum
  threshold, we find that the Efimov state in medium can turn into a
  quasibound state with a finite decay width.
\end{abstract}
\maketitle


\section{Introduction}

The interaction of impurity particles in a medium is studied across
physical disciplines.  Specifically, the Casimir interaction between
two impurities arises from fluctuations of the medium, or even the
vacuum, subject to the boundary conditions imposed by the impurities
\cite{casimir1948}.  Current applications range from neutron stars
\cite{yu2000} and the quark-gluon plasma \cite{neergaard2000} to
ultracold atoms \cite{nishida2009casimir, macneill2011}.  Recent
advances in experiments with ultracold atomic gases allow exploring
mobile impurities in a fermionic medium, or Fermi polarons, in the
regime of strong attraction \cite{schirotzek2009, nascimbene2009,
  kohstall2012, koschorreck2012, cetina2016, scazza2017,
  yan2019boiling} and precisely measuring their spectral properties.
These experiments are performed not on a single impurity but on a
dilute gas of impurities.  The induced interaction between impurities
is typically weak \cite{mora2010, desalvo2019, edri2020,
  mukherjee2020}, but it can play an important role when the
impurity-medium interaction becomes strong.  Indeed, for large
scattering length it can lead to Efimov three-body bound states
\cite{efimov1970, braaten2006, naidon2017} that are crucial for
interpreting impurity spectra \cite{lompe2010}.

The interaction between localized spins in an electron gas is a
classic result of condensed matter physics: by the Pauli principle,
the induced Ruderman-Kittel-Kasuya-Yosida (RKKY) interaction
oscillates in space and changes sign whenever the distance between the
spins grows by about an electron spacing, or Fermi wavelength
\cite{ruderman1954, kasuya1956, yosida1957}.  For larger objects in a
Fermi sea, this can be understood by semiclassical methods
\cite{bulgac2001}.  More recently, these studies have been extended to
the case of impurity atoms in a Fermi gas, or Fermi polarons
\cite{chevy2006, schmidt2011, massignan2014}.  When the impurity is
tuned to strong attraction with the Fermi sea, it can form a bound
state with one of the fermions \cite{schmidt2011, punk2009, mora2009}.
These, in turn, lead to an enhanced attraction between two impurities
at short distance \cite{nishida2009casimir, macneill2011, endo2013}
and even to bipolaron bound states between two impurities in a Fermi
sea \cite{macneill2011, nygaard2014, bellotti2016, sun2019efimov,
  tran2020}.  The Efimov bound states between two impurities and one
fermion are characterized by discrete scaling relations
\cite{braaten2006, ulmanis2016}.  In the medium, the scaling relations
are modified by the Fermi wavelength as an additional length scale
\cite{macneill2011, nygaard2014, sun2019efimov, tran2020} and lead to
shifts in the bipolaron resonance positions.  Because they satisfy a
new scaling relation, we shall refer to them as in-medium Efimov
resonances.  In the limit of a dense medium the induced interaction
diminishes proportional to the Fermi wavelength and eventually
vanishes \cite{endo2013}.

\begin{figure}[b]
  \centering
  \includegraphics[width=.8\linewidth]{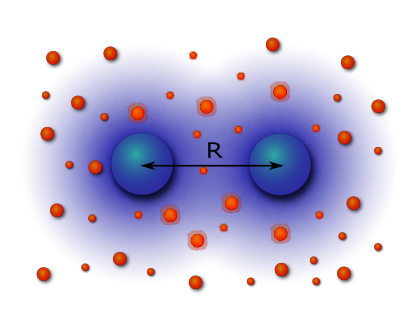}
  \caption{Two heavy impurities (large blue dots) at distance $R$ in a
    Fermi sea of light fermions (small red dots).}
  \label{fig:sketch}
\end{figure}

In this work, we study the scattering properties of two heavy
impurities in an ideal Fermi gas, as shown in Fig.~\ref{fig:sketch}.
Based on the Casimir interaction potential \cite{nishida2009casimir},
we compute the scattering phase shift and the induced scattering
length between impurities and find that they scatter resonantly
whenever an Efimov bound state appears at the continuum threshold.
Moreover, for positive scattering length a repulsive barrier arises in
the impurity potential, and remarkably the in-medium Efimov state can
live on behind the barrier as a quasibound state at positive
energies.  In the following, we start by reviewing the Casimir
interaction potential in Sec.~\ref{sec:pot}.  In
Section~\ref{sec:scat} we solve the Schr\"odinger equation for two
impurities in this potential to find the scattering properties, and we
discuss our results also for the experimentally relevant case of
cesium-lithium mixtures \cite{pires2014, tung2014} before concluding
in Sec.~\ref{sec:conc}.


\section{Casimir interaction}
\label{sec:pot}

The interaction of two heavy impurities (mass $M$) in an ideal Fermi
gas of light particles (mass $m$) is well described in the
Born-Oppenheimer approximation.  By the separation of time scales, the
impurities can be considered as a static scattering potential for the
fermions and---in the case of a contact potential---provide only a
boundary condition for the fermion wavefunctions.  This approximation
becomes exact in the limit of infinitely heavy impurities, where the
problem reduces to potential scattering, and remains accurate at large
mass ratio $M/m\gg1$, for instance in a quantum gas mixtures of
bosonic $^{133}$Cs and fermionic $^6$Li atoms.  In this section we
present the derivation of the interaction $V(R)$ induced between the
two heavy impurities (of arbitrary statistics) by the presence of the
Fermi sea, following Nishida \cite{nishida2009casimir}.

Consider two infinitely heavy impurities at distance $R$ with
positions $\vec R_{1,2} = \pm \vec R/2$.  The impurities have a
short-range attractive interaction with the fermions, which we model
by a zero-range Fermi pseudopotential.  The action of the potential is
equivalent to imposing the Bethe-Peierls boundary condition on the
fermion wavefunction near an impurity at position $\vec R_i$,
\begin{align}
  \label{eq:bc}
  \psi(\vec x \to \vec R_i) \propto \frac1{\abs{\vec x-\vec R_i}} -
  \frac1a + \mathcal O(\abs{\vec x-\vec R_i}).
\end{align}
Here, $a$ denotes the impurity-fermion scattering length that fully
characterizes the contact interaction.  The fermion wavefunctions
solve the free Schr\"odinger equation, subject to the boundary
conditions \eqref{eq:bc} at both $\vec R_1$ and $\vec R_2$.  There are
potentially two bound states at negative energies
$E_\pm=-\kappa_\pm^2/2m<0$, where the inverse length scale of the
bound states $\kappa_\pm>0$ is given by
\begin{align}
  \kappa_\pm = \frac1a + \frac1R W(\pm e^{-R/a})
\end{align}
in terms of the Lambert W function that solves $x=W(x)e^{W(x)}$.
Since real solutions exist for $x\in(-1/e,\infty)$, the bound state
$\kappa_\pm>0$ appears for distances $R/a>\mp1$: while $\kappa_-$
exists only for positive scattering length and $R>a>0$, $\kappa_+$
exists both for $a<0$ at small separation $R<\abs a$ and for $a>0$ at
arbitrary $R$.  Hence, a fermion attracted to two impurities forms a
$\kappa_+$ bound state much more \emph{easily} than one attracted only
to a single impurity, and this will have dramatic consequences for the
scattering properties between two impurities.

\begin{figure}[t]
  \centering
  \includegraphics[width=\linewidth]{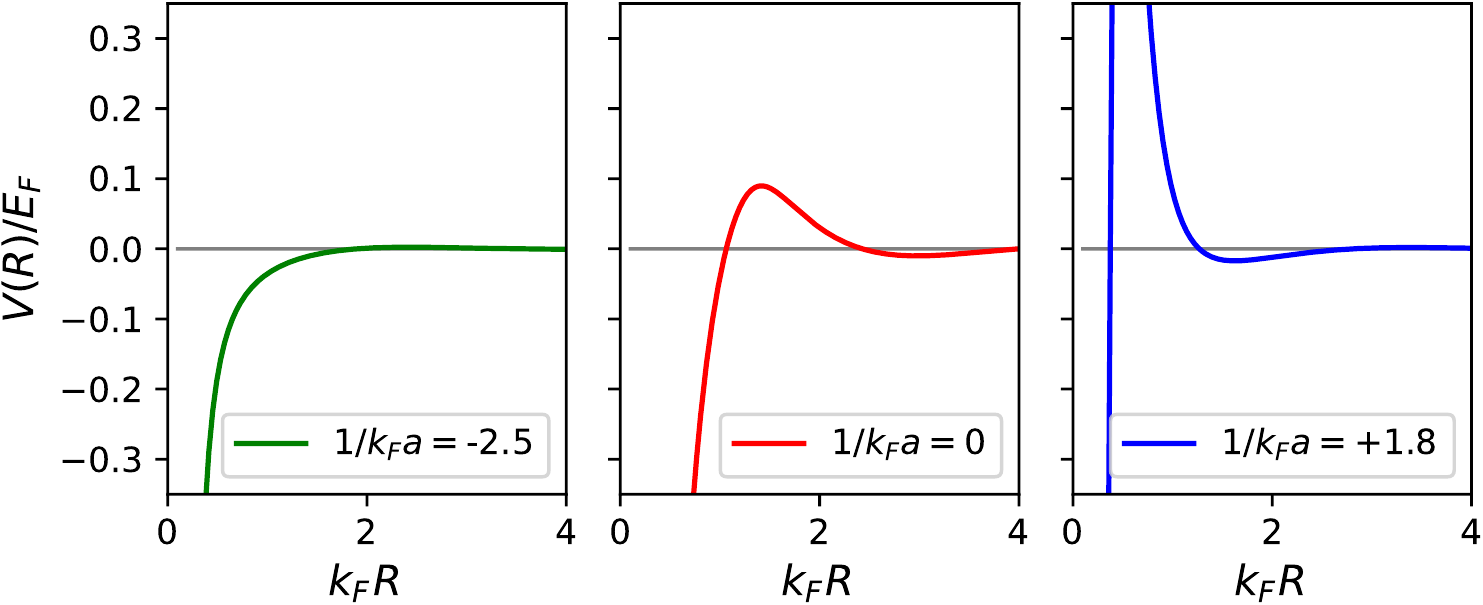}
  \caption{Induced interaction potential $V(R)$ between two heavy
    impurities for negative, unitary, and positive interspecies
    scattering length $a$ (from left to right).  Data shown for Cs-Li
    mass ratio $M/m=22.17$.}
  \label{fig:potential}
\end{figure}

Besides the bound states, there is a continuum of fermion scattering
states at positive energy $E=k^2/2m>0$.  For each mode $\vec k$, the
fermion wavefunction $\sin(kr+\delta_\pm)$ at large distance $r$ from
both impurities acquires an $s$-wave phase shift with respect to the
free wavefunction without impurities, which is given by
\begin{align}
  \tan \delta_\pm(k) = -\frac{kR\pm\sin(kR)}{R/a\pm\cos(kR)}
\end{align}
for the (anti-)symmetric solution, where $0\leq \delta_\pm(k) < \pi$.
In the thermodynamic limit, the total energy change with and without
impurities can be expressed as
\begin{align}
  \label{eq:DelE}
  \Delta E(R) = -\frac{\kappa_+^2+\kappa_-^2}{2m} - \int_0^{k_F} dk\,
  k\frac{\delta_+(k)+\delta_-(k)}{\pi m}.
\end{align}
At large separation the impurities no longer interact, and the energy
change approaches
\begin{align}
  \Delta E(R\to\infty) \to 2\mu,
\end{align}
or twice the single-polaron energy (chemical potential)
\begin{align}
  \mu = -\ef \frac{k_Fa +
  [1+(k_Fa)^2][\pi/2+\arctan(1/k_Fa)]} {\pi(k_Fa)^2}
\end{align}
in terms of the Fermi energy $\ef=k_F^2/2m$.  The resulting Casimir
interaction relative to the chemical potential,
\begin{align}
  \label{eq:VR}
  V(R) = \Delta E(R) - 2\mu,
\end{align}
is shown in Fig.~\ref{fig:potential}.  For short distance it is
strongly attractive as $-c^2/2mR^2$ from the bound-state contribution
$\kappa_+$, where $c=W(1)\approx0.567\,143$ solves $c=e^{-c}$; this
effect is already present for a single fermion and gives rise to the
Efimov effect \cite{efimov1970, braaten2006, naidon2017}.  In the
fermionic medium, the Pauli principle requires that the induced
interaction changes sign after an average spacing between the
fermions, similar to the RKKY interaction in solids
\cite{ruderman1954, kasuya1956, yosida1957}.  The strong attraction is
thus cancelled at larger distances by the contribution from the Fermi
sea and crosses over near $k_FR\simeq1$ into an oscillating decay
$\cos(2k_FR)/R^3$ at large distance.  Specifically at unitarity, the
bound-state contribution $-c^2/2mR^2$ is present for \emph{all} $R$
and is cancelled by the Fermi-sea contribution
$2\mu+c^2/2mR^2-\cos(2k_FR)/2\pi mk_FR^3 + \mathcal O((k_FR)^{-4})$.
For positive scattering length, a substantial repulsive barrier
develops that will be able to capture a quasibound state, as we will
discuss in the next section.


\section{Scattering between impurities}
\label{sec:scat}

Given the induced potential $V(R)$ between the impurities, we now
generalize the approach of Ref.~\cite{nishida2009casimir} to bosonic
or distinguishable impurities and compute their scattering properties
in the $s$-wave channel.  We still work in the Born-Oppenheimer
approximation where the heavy impurities move slowly, while the Fermi
sea of light particles adjusts almost instantaneously to their
positions and produces the potential.  The stationary states of the
impurities are then described by the Schr\"odinger equation
\begin{align}
  \label{eq:schr}
  \Bigl[ -\frac{\nabla_{\vec R}^2}M + V(R) + 2\mu - E\Bigr]
  \Psi(\vec R) = 0
\end{align}
in the central potential $V(R)$.  The scattering properties are
encoded in the scattering phase shifts $\delta_\ell^\text{ind}(k)$
induced by the medium in the $\ell$ partial wave component.  We
compute the $s$-wave phase shift by integrating the variable phase
equation \cite{calogero1967}
\begin{align}
  \label{eq:var}
  k\partial_R \delta_{\ell=0}^\text{ind}(k,R)
  = -MV(R) \sin[kR+\delta_{\ell=0}^\text{ind}(k,R)]^2.
\end{align}
Usually, one imposes the boundary condition
$\delta_{\ell=0}^\text{ind}(k,R=0)=0$ at $R=0$ and integrates up to
large $R$, where one reads off the phase shift
$\delta_{\ell=0}^\text{ind}(k) =
\delta_{\ell=0}^\text{ind}(k,R\to\infty)$.

\subsection{Efimov resonances}

The short-range singularity of the induced potential
$V(R\to0)=-\alpha/R^2$ leads to a Hamiltonian that is bounded from
below only for weak attraction $\alpha<1/4$; for larger $\alpha$ there
are an infinite number of Efimov bound states \cite{braaten2006}.  In
our case $\alpha=(M/2m)c^2$ is always above $1/4$ in the
Born-Oppenheimer limit $M\gg m$, so the potential needs a
regularization, which is physically provided by the repulsive core of
the van der Waals potential between impurities \cite{tran2020}.  We
mimic the actual potential by a hard sphere of radius $R_0$, where the
initial condition reads $\delta_{\ell=0}^\text{ind}(k,R_0)=-kR_0$, and
integrate $R=R_0\dotsc\infty$ using a standard ODE solver (DOP853).
The cutoff radius $R_0$ is tuned to match the size of the lowest
Efimov state in the real potential and is therefore directly related
to the three-body parameter (3BP) which incorporates the relevant
short-range physics \cite{braaten2006, wang2012}.  As a specific
example, in the Cs-Li system the heteronuclear Feshbach resonance at
$889\,$G has $a_-^{(1)}=-2130\,a_B$ \cite{haefner2017}, which is
reproduced by the induced potential with $R_0=220\,a_B$.  For a
typical fermion density of $n=10^{13}\,\text{cm}^{-3}$ in current
experiments \cite{cetina2016, edri2020} we thus obtain $k_FR_0=0.1$
and we use this value in our plots to make quantitative predictions.

\begin{figure}[t]
  \centering
  \includegraphics[width=\linewidth]{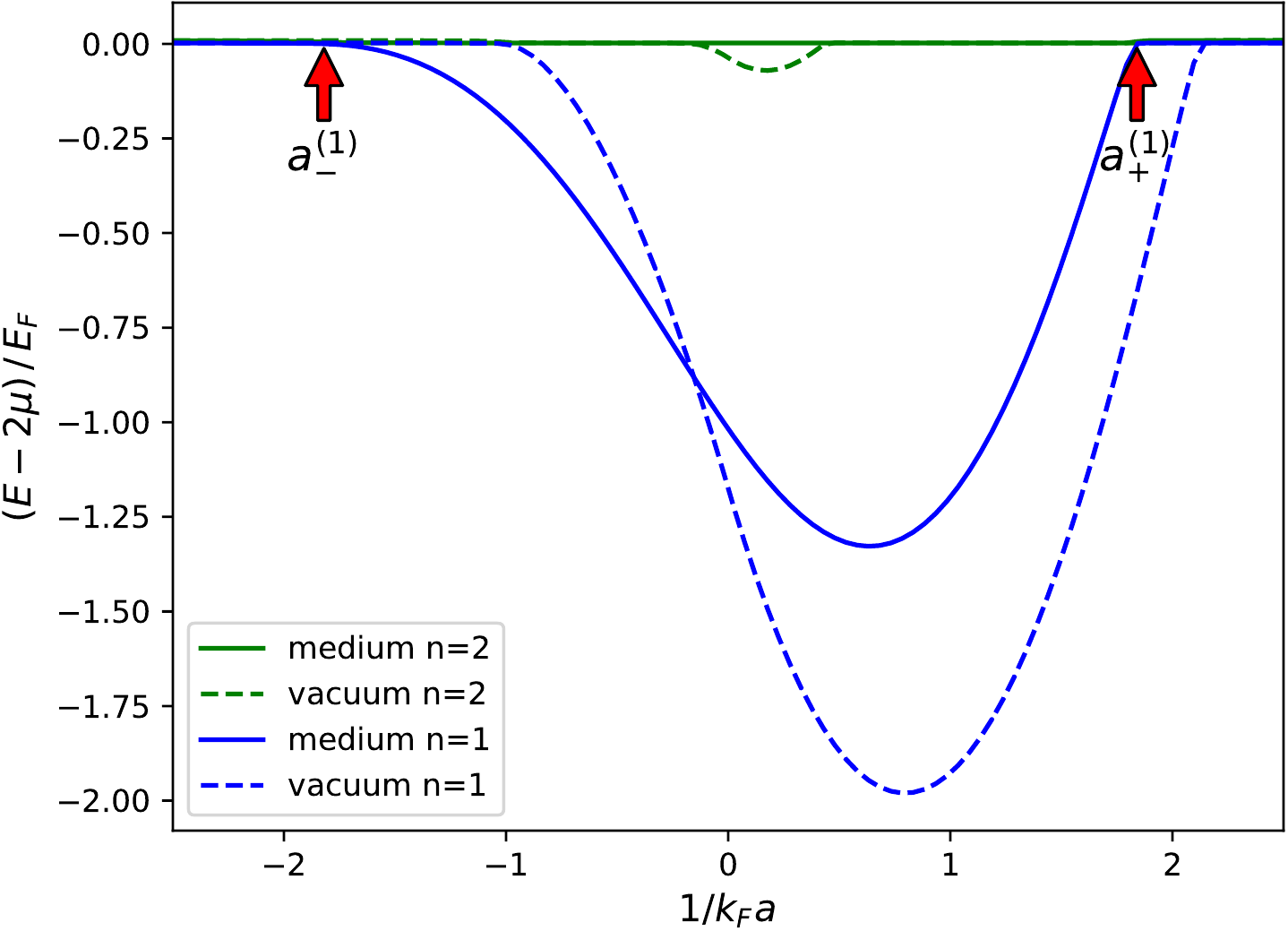}
  \caption{Energy spectrum of Cs-Cs-Li Efimov states vs
    impurity-fermion scattering.  The energies are given relative to
    the scattering continuum $2\mu$.  Shown are the first ($n=1$, blue
    (lower) lines) and second ($n=2$, green (upper) lines) Efimov
    states, both in vacuum (dashed) and in medium (solid), with cutoff
    radius $R_0=0.1\,k_F^{-1}$.  The Efimov bound states merge with
    the continuum at scattering lengths $a_\pm^{(n)}$, as indicated by
    the arrows for the first in-medium Efimov state.  In vacuum,
    length units are $10R_0$ and energy units $1/2m(10R_0)^2$.}
  \label{fig:specrel}
\end{figure}

The bound-state spectrum for Eq.~\eqref{eq:schr} is shown in
Fig.~\ref{fig:specrel} for the example of $^{133}$Cs impurities in a
$^6$Li Fermi sea.  One observes that the medium facilitates binding
for weak attraction (shifting the onset to the left), but the
repulsive barrier inhibits binding compared to the vacuum case for
strong attraction \cite{sun2019efimov}.

\subsection{Induced scattering length}

\begin{figure}[t]
  \centering
  \includegraphics[width=\linewidth]{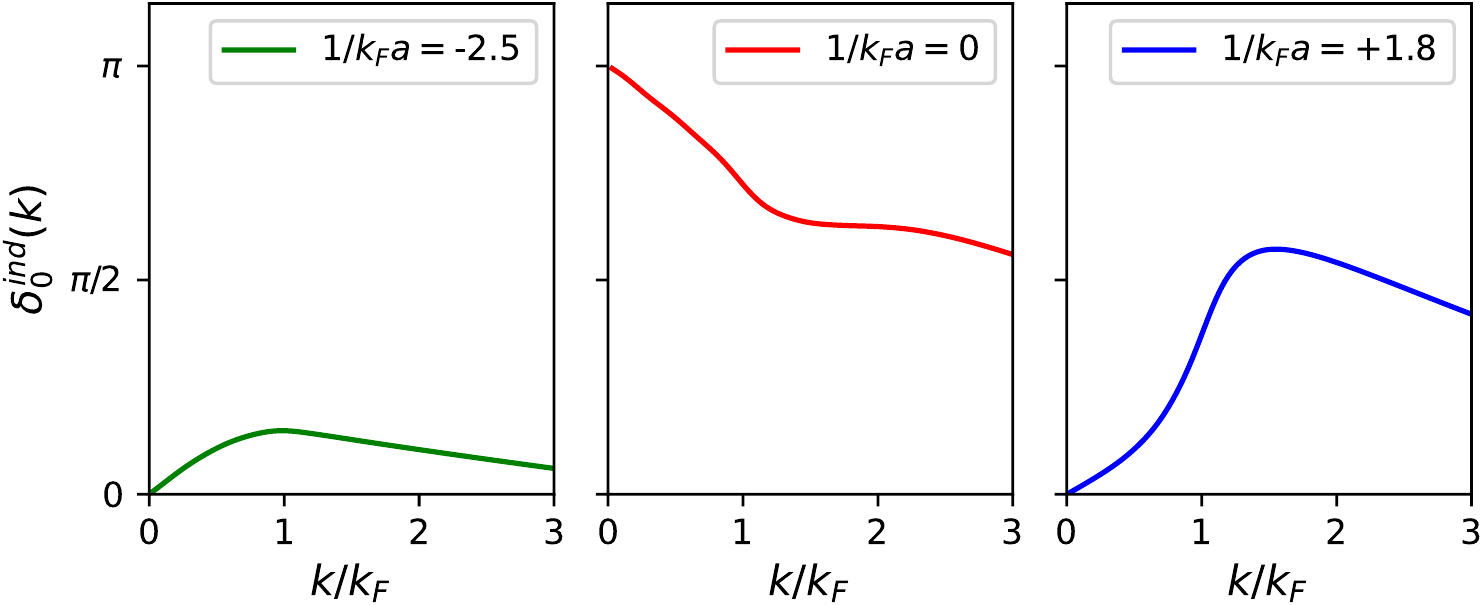}
  \caption{Induced scattering phase shift $\delta_0^\text{ind}(k)$ for
    the potentials in Fig.~\ref{fig:potential}.  The initial slope
    near $k=0$ determines the induced scattering lengths
    $k_Fa_\text{ind} = -0.8; +0.7; -1.0$ from left to right.  On the
    $a>0$ side the phase shift steeply rises above $\pi/2$ indicating
    a quasibound state.  Data shown for Cs-Li mass ratio
    $M/m=22.17$.}
  \label{fig:phaseshift}
\end{figure}

For a given cutoff radius $R_0$ and the corresponding Efimov spectrum,
we compute the resulting $s$-wave scattering phase shifts
$\delta_{\ell=0}^\text{ind}(k)$ that are shown in
Fig.~\ref{fig:phaseshift}.  In the limit of small $k$ one can read
off the induced impurity-impurity scattering length $a_\text{ind}$
shown in the figure and the effective range $r_\text{e}$ from the
effective range expansion
\begin{align}
  \label{eq:aind}
  k\cot[\delta_{\ell=0}^\text{ind}(k)]
  = -\frac1{a_\text{ind}} + \frac{r_\text{e}}2 k^2 + \mathcal O(k^4).
\end{align}
Equivalently, the scattering length can be obtained from the variable
phase equation \eqref{eq:var} directly in the $k\to0$ limit,
\begin{align}
  \label{eq:vara}
  \partial_R a_\text{ind}(R) = -MV(R) [R-a_\text{ind}(R)]^2,
\end{align}
with initial condition $a_\text{ind}(R_0)=R_0$ and the final result
$a_\text{ind} = a_\text{ind}(R\to\infty)$.  The Efimov bound states
lead to resonances in the induced scattering length \cite{endo2011},
which are understood analytically from the solution of
Eq.~\eqref{eq:vara} for the $-\alpha/R^2$ potential with $\alpha>1/4$
for distances $R_0\dotsc R$,
\begin{align}
  a_\text{ind}(R) = R\Bigl[ 1-\frac1{2\alpha} +
  \frac{s_0}\alpha \tan \Bigl(
  \arctan\frac1{2s_0} - s_0 \ln \frac
  R{R_0}\Bigr) \Bigr]
\end{align}
with $s_0=\sqrt{\alpha-1/4}>0$ \cite{moroz2011}.  This solution is
valid for distances $R_0 < R\ll \abs a,k_F^{-1}$ and shows that the
continuous scale invariance of the $1/R^2$ potential is broken down to
a discrete scaling symmetry.  The solution repeats itself whenever
$s_0 \ln (R/R_0)$ is a multiple of $\pi$, hence is log-periodic in $R$
with a length scale factor of $l=\exp(\pi/s_0)$.  For the case of
$^{133}$Cs impurities in $^6$Li, the scale factor is $l\approx 5.6$ in
the Born-Oppenheimer approximation, close to the experimentally
observed value of $l\approx4.9$ \cite{ulmanis2016}.  For larger
distance $R \gtrsim k_F^{-1}$ the $-\alpha/R^2$ form of the potential
is cut off by the Fermi sea, and no Efimov bound states of size larger
than $k_F^{-1}$ occur.

\begin{figure}[t]
  \centering
  \includegraphics[width=\linewidth]{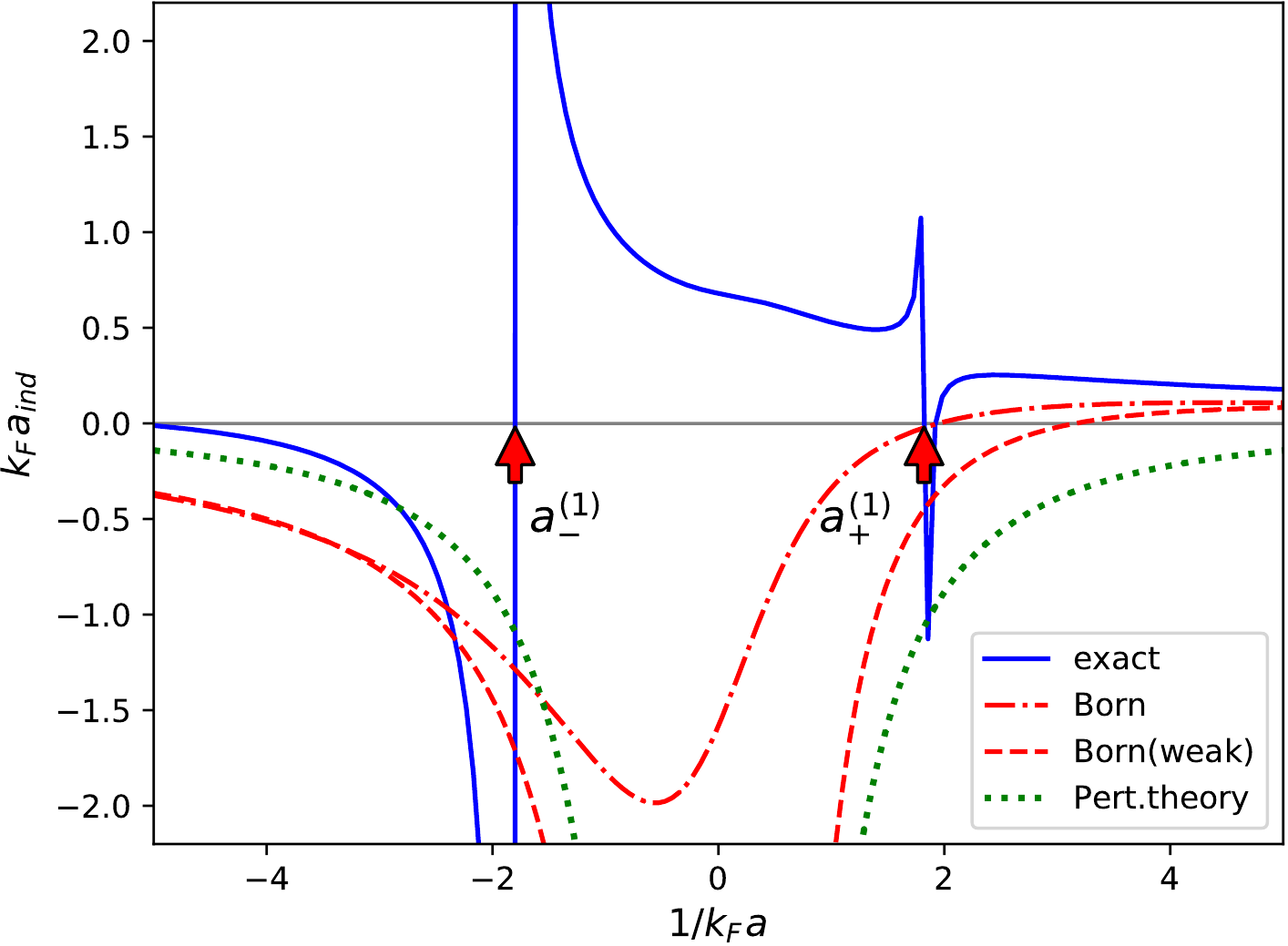}
  \caption{Induced interaction between two heavy impurities: induced
    scattering length $a_\text{ind}$ vs impurity-fermion interaction.
    From top to bottom: exact solution of Schr\"odinger equation
    \eqref{eq:vara} with cutoff $k_FR_0=0.1$ (blue solid line), Born
    approximation \eqref{eq:born} (red dash-dotted line), analytical
    weak-coupling Born approximation \eqref{eq:bornweak} (red dashed
    line), and second-order perturbation theory \eqref{eq:aindPT}
    (green dotted line).  The exact $a_\text{ind}$ diverges at the
    in-medium Efimov resonances $a_\pm^{(1)}$ indicated by the arrows.
    Data shown for Cs-Li mass ratio $M/m=22.17$.}
  \label{fig:aind}
\end{figure}

The full potential $V(R)$ in Eqs.~\eqref{eq:DelE}, \eqref{eq:VR} is
computed numerically and agrees with known analytical limits for small
or large distance and weak or strong coupling
\cite{nishida2009casimir}.  The induced scattering length for the full
potential is shown in Fig.~\ref{fig:aind} for $k_FR_0=0.1$ (blue solid
line).  In this case, $a_\text{ind}$ exhibits two scattering
resonances at $a=a_\pm^{(1)}$, where a bound state crosses the
continuum threshold.  For smaller $R_0$, the potential admits more
bound states and associated resonances at $a=a_\pm^{(n)}$ with $n>1$,
for comparison see Fig.~2(b) in Ref.~\cite{sun2019efimov}.  In the
interval $1/a_-^{(n)}<1/a<1/a_+^{(n)}$ the induced potential admits
$n$ Efimov bound states, and the phase shift starts at
$\delta_{\ell=0}^\text{ind}(k\to0) = n\pi$ in accordance with
Levinson's theorem, as shown for $n=1$ in the central panel of
Fig.~\ref{fig:phaseshift}.

The resonances of $a_\text{ind}(a)$ occur whenever an Efimov bound
state crosses the continuum threshold.  This can be seen in the energy
spectrum in Fig.~\ref{fig:specrel}: for $1/a>1/a_-^{(n)}$ the
potential is deep enough to admit the $n$th bound state, but for even
stronger attraction this bound state eventually merges again with the
scattering continuum at $1/a=1/a_+^{(n)}$.  Note that the resonance
positions $a_\pm^{(n)}(k_F)$ in medium depend on the density and
differ from the vacuum values $a_\pm^{(n)}(0)$, as discussed in
Refs.~\cite{nygaard2014, sun2019efimov, tran2020}.

For the singular potential $V(R)$, the exact induced scattering length
$a_\text{ind}$ can differ drastically from the one obtained in Born
approximation,
\begin{align}
  \label{eq:born}
  a_\text{ind}^\text{Born} = \int_0^\infty dR\, R^2\, MV(R).
\end{align}
Here, the asymptotics at short distance [$1/R^2$] and at large
distance [$\cos(2k_FR)/R^3$] are integrable and no cutoff $R_0$ is
needed.  The resulting scattering length is shown in
Fig.~\ref{fig:aind} (red dash-dotted line); as might be expected for a
singular potential, it does not approximate the exact solution well
even for weak coupling.

It is instructive to compare the induced scattering length to the
exact result in the weakly attractive limit $1/k_Fa\lesssim-1$.  In
this case, the full induced potential is given analytically for all
$R$ as the sum of the singular attractive potential from the bound
state and the regular oscillating potential from the Fermi sea,
\begin{multline}
  V_\text{weak}(R) = -\frac{\Theta(\abs a-R)}{2mR^2} \Bigl( W(e^{R/\abs
    a}) - \frac R{\abs a} \Bigr)^2 \\
  + \frac{a^2}{2m}\, \frac{2k_FR
  \cos(2k_FR) - \sin(2k_FR)}{2\pi R^4} + \mathcal O\bigl((k_Fa)^3\bigr).
\end{multline}
For weak coupling, we find an analytical expression for the induced
scattering length in Born approximation with $R_0=0$ [red dashed line
in Fig.~\ref{fig:aind}],
\begin{align}
  \label{eq:bornweak}
  a_\text{weak}^\text{Born} = \frac M{2m} \Bigl( \gamma a -
  \frac{k_F}\pi a^2 + \mathcal O(a^3) \Bigl)
\end{align}
where
$\gamma = \int_0^1 dx [W(e^x)-x]^2 = 2(1-c[1+c(1+c/3)]) \approx
0.100\,795$.  Figure~\ref{fig:aind} shows that the analytical
weak-coupling form \eqref{eq:bornweak} agrees with the numerical Born
solution \eqref{eq:born} for $\abs{k_Fa}\lesssim 0.3$.  Finally,
second-order perturbation theory for weakly repulsive interaction
yields \cite{santamore2008}
\begin{align}
  \label{eq:aindPT}
  a_\text{ind}^\text{PT} = -\frac{k_F}{2\pi}\, \frac{(M+m)^2}{Mm}\,
  a^2 + \mathcal O(a^3)
\end{align}
from the continuum of scattering states alone [green dotted line in
Fig.~\ref{fig:aind}].  This result at order $\mathcal O(a^2)$ fully
agrees with the second-order term in the Born approximation
\eqref{eq:bornweak} in the Born-Oppenheimer limit $M\gg m$.  However,
the first term in the Born approximation \eqref{eq:bornweak} that
arises from the bound state is of \emph{first order} in $a$ and
therefore dominates over the continuum contribution at weak coupling
$R_0 < \abs a \lesssim k_F^{-1}$.  Hence, the usual perturbation
theory for repulsive impurities is unable to describe attractive
impurities even at weak coupling because it misses the leading
bound-state contribution for $\abs a>R_0$.  In the exact solution of
the Schr\"odinger equation, the bound-state contribution can become
arbitrarily large near an Efimov resonance, depending on the value of
the cutoff radius $R_0$.  Only for very weak attraction with
$\abs a\lesssim R_0$ the bound-state contribution is small, and the
induced scattering is dominated by the second-order contribution
\eqref{eq:aindPT}, as is the case in Ref.~\cite{desalvo2019} where
$k_Fa\approx -0.012$, and in Ref.~\cite{edri2020}.

\subsection{quasibound states}

\begin{figure}[t]
  \centering
  \includegraphics[width=\linewidth]{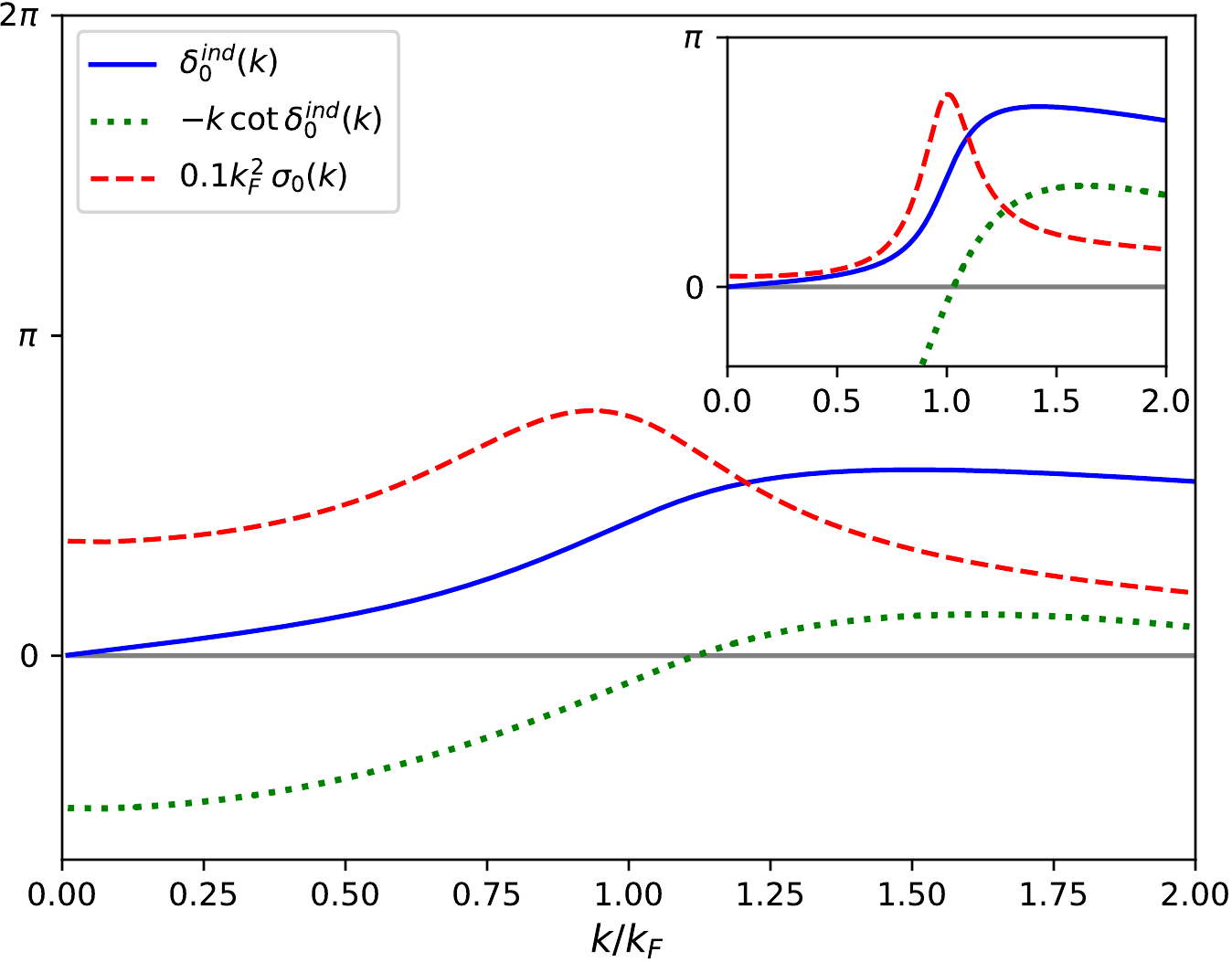}
  \caption{Scattering resonance at positive energy above the continuum
    threshold and quasibound states.  Enhanced scattering occurs at
    the upturn of the scattering phase shift $\delta_0^\text{ind}(k)$
    (blue solid line) at $k=1.12\,k_F$, where
    $\cot\delta_0^\text{ind}(k)$ has a zero crossing (green dotted
    line).  Correspondingly, the $s$-wave differential cross section
    $\sigma_0$ (red dashed line) exhibits a maximum at energies near
    $\ef$.  The data shown are for the Cs-Li mass ratio $M/m=22.17$
    and $k_FR_0=0.1$ near the Efimov resonance at
    $a=0.536\,k_F^{-1} \lesssim a_+^{(1)} = 0.542\,k_F^{-1}$.  Inset:
    for larger mass ratio $M/m=44.33$ there is a well-developed
    scattering resonance of Breit-Wigner form \eqref{eq:qbound} with
    $E_\text{qbnd}=1.06\,\ef$ and $\Gamma_\text{qbnd}=0.69\,\ef$.
    This arises from a quasibound state at
    $a=0.359\,k_F^{-1} \lesssim a_+^{(1)} = 0.360\,k_F^{-1}$.}
  \label{fig:quasibound}
\end{figure}

Beyond the Efimov threshold $1/a>1/a_+^{(n)}$ at positive scattering
length, the in-medium Efimov bound state is pushed out of the
potential to energies above the continuum treshold, but it may be
caught behind the repulsive barrier that is created by the fermionic
medium and the two-body bound states [right panel of
Fig.~\ref{fig:potential}].  How long the bound state can be caught
behind the barrier depends on the effective height of the potential in
the Schr\"odinger equation \eqref{eq:schr}, which is proportional to
the mass ratio $M/m$.  The larger the mass ratio, the longer lived is
the quasibound state even at positive energies.  We find long-lived
states approximately for $M/m \gtrsim 40$.  In this case, the Efimov
bound state goes over into a quasibound state at positive energies
and with a small decay width, similar to the collisionally stable
quasibound states found in Ref.~\cite{kartavtsev2007}.  We identify
such a state when the scattering phase shift assumes the form of a
Breit-Wigner resonance at positive energies $E=k^2/2m$ as shown in
Fig.~\ref{fig:quasibound},
\begin{align}
  \label{eq:qbound}
  \cot\bigl[\delta_{\ell=0}^\text{ind}(k)\bigr]
  = -\frac{E-E_\text{qbnd}}{\Gamma_\text{qbnd}/2} + \dotsm
\end{align}
From the position of the zero crossing and the slope we read off the
energy $E_\text{qbnd}$ and the FWHM decay width $\Gamma_\text{qbnd}$.
For Cs-Li parameters $M/m=22.17$, Fig.~\ref{fig:quasibound} shows
enhanced scattering at positive energies but still large width
$\Gamma_\text{qbnd} > E_\text{qbnd}$ so that we cannot yet speak of a
well-defined quasibound state.  For larger mass ratio $M/m=44.33$, we
find that for $1/a>1/a_+$ the in-medium Efimov state can turn into a
well-developed quasibound state as shown in the inset: it has a decay
width $\Gamma_\text{qbnd} < E_\text{qbnd}$ smaller than its energy.

Based on \cite{naidon2017, sun2019efimov}, it appears reasonable to
assume that the excited quasibound trimer state will eventually decay
into two polarons, which form the continuum threshold for $1/a>1/a_+$.
The character of these polaron states depends on the scattering length
across the polaron-to-molecule transition \cite{punk2009, mora2009,
  schmidt2011}.  For strong binding $1/k_Fa > (1/k_Fa)_c \simeq 0.9$,
which is the situation depicted in Fig.~\ref{fig:specrel} near
$a_+^{(1)}$, each impurity forms a tightly bound impurity-fermion
dimer of energy $\mu$ embedded in the residual Fermi sea
\cite{naidon2017, sun2019efimov}.  For weaker binding
$1/k_Fa < (1/k_Fa)_c$, instead, each impurity forms a Fermi polaron,
which would describe the continuum threshold near higher-lying Efimov
states $a_+^{(n>1)}$.

A quasibound state is also manifest as a peak in the $s$-wave
scattering cross section [red dashed line in
Fig.~\ref{fig:quasibound}]
\begin{align}
  \sigma_{\ell=0}(k)
  = \frac{4\pi}{k^2} \sin^2\bigl[\delta_{\ell=0}^\text{ind}(k)\bigr]
\end{align}
at positive energy.  For a finite density of heavy impurities in
thermal equilibrium with the medium at $T \simeq E_\text{qbnd}$ there
will be enhanced scattering between the impurities, which would lead
to a greater mean-field shift in the impurity spectra proportional to
the impurity density.

Experimentally, the Efimov bound states in medium could be observed as
a medium-density dependent shift of the three-body loss peaks
associated with the Efimov trimers \cite{sun2019efimov}.  The
quasibound state and scattering resonance at positive energies above
the continuum threshold would lead to an impurity-density dependent
shift in the polaron spectrum, estimated at a few percent in the case
of Ref.~\cite{naidon2018}, and to enhanced radiofrequency association
of Efimov trimers \cite{lompe2010} beyond $a_+^{(n)}$.


\section{Conclusion}
\label{sec:conc}

The induced interaction between \emph{attractive} impurities in a
Fermi sea differs fundamentally from the RKKY interaction between
nuclear spins in an electron gas, or repulsive impurities.  While the
continuum of scattering states yields a similar oscillating potential
at large distance, the appearance of bound states implies a strong
attraction at short distances.  This singular $-1/R^2$ attraction
gives rise to a series of three-body Efimov bound states down to the
cutoff scale.  Whenever a bound state crosses the continuum threshold,
the induced scattering length $a_\text{ind}$ exhibits resonances and
changes sign.  Attractive impurities can thus scatter strongly, and
repulsively, in distinction to the weak induced attraction for
repulsive impurities.  For very weak attraction of order
$k_Fa\approx-0.01$, instead, our prediction for the induced scattering
length is just slightly more attractive than in perturbation theory
due to the additional attraction by the bound state, consistent with
recent measurements \cite{desalvo2019, edri2020}.

While the impurity-impurity-fermion Efimov bound states below the
continuum threshold have been discussed earlier, we find that at
positive scattering length and large mass ratio the Efimov states can
turn into quasibound states at positive energy.  This corresponds to
two impurities caught behind the repulsive potential barrier created
by the Fermi sea: they can eventually tunnel through the barrier and
escape, but as long as they are close, there is an enhanced
probability to form a deeply bound state.  This three-body
recombination leads to a clear signature in experimental loss spectra
\cite{pires2014, tung2014, ulmanis2016}.

Our investigation can be generalized to a dilute gas of heavy
impurities, where it has been shown that the total Casimir energy is
well approximated by a sum of pairwise two-body energies
\cite{bulgac2001, nishida2009casimir}.  It is then justified to apply
our results to a thermal gas of impurities at temperature $T$, where
the scattering properties are evaluated at the thermal wavevector
$\lambda_T^{-1} = \sqrt{mT/2\pi}$.  This leads to the prediction of an
enhanced mean-field shift when $T\simeq E_\text{qbnd}$.  Furthermore,
if three impurities are all nearby it would be interesting to explore
the emergence of four-body impurity-impurity-impurity-fermion bound
states.  For smaller mass ratio, corrections beyond the
Born-Oppenheimer approximation have to be included \cite{endo2013,
  schecter2014}, in particular the scattering of trimers by the Fermi
sea, which creates particle-hole excitations and alters the induced
potential \cite{macneill2011}.

For the related case of impurities in a Bose-Einstein condensate,
recent studies found many-body bound states of two impurities, or
bipolarons, for moderately attractive interaction \cite{zinner2013,
  naidon2018, camacho2018bipolarons, camacho2018landau}.  It will be
interesting to extend these studies to the regime of strong attraction
on the molecular side of the Feshbach resonance, where the impurities
have been shown to strongly deform the surrounding BEC
\cite{drescher2019, drescher2020exact}.  This again gives rise to an
oscillating induced potential between the impurities that can be
described using nonlocal Gross-Pitaevskii theory \cite{drescher2020}.

\begin{acknowledgments}
  We thank A. Volosniev for interesting discussions.  This work is
  supported by the Deutsche Forschungsgemeinschaft (DFG, German
  Research Foundation), project-ID 273811115 (SFB1225 ISOQUANT) and
  under Germany's Excellence Strategy EXC2181/1-390900948 (the
  Heidelberg STRUCTURES Excellence Cluster).  E.L.\ acknowledges
  support by the IMPRS-QD.
\end{acknowledgments}


\bibliography{all}

\end{document}